%% file: main.tex
\documentclass[preprint]{article}
\usepackage{amsmath,graphicx}
\usepackage{booktabs}
\usepackage{multirow}
\usepackage[preprint]{spconf}
\newfont{\mycrnotice}{ptmr8t at 7pt}
\newfont{\myconfname}{ptmri8t at 7pt}
\title{ASR4REAL: an extended benchmark for speech models}
\name{Morgane Riviere$^\dagger{}$, Jade Copet$^\dagger{}$, Gabriel Synnaeve$^\dagger{}$}
\address{$\dagger{}$Facebook AI Research}
\toappear{Submitted to ICASSP 2022}
\begin{document}
\input{macros}
%
\maketitle
\begin{abstract}
Popular ASR benchmarks such as Librispeech and Switchboard are limited in the diversity of settings and speakers they represent.
We introduce a set of benchmarks matching real-life conditions, aimed at spotting possible biases and weaknesses in models. We have found out that even though recent models do not seem to exhibit a gender bias, they usually show important performance discrepancies by accent, and even more important ones depending on the socio-economic status of the speakers. Finally, all tested models show a strong performance drop when tested on conversational speech, and in this precise context even a language model trained on a dataset as big as Common Crawl does not seem to have significant positive effect which reiterates the importance of developing conversational language models.
\end{abstract}

\begin{keywords}
Speech-to-Text, Benchmark, Robustness
\end{keywords}
\section{Introduction}
\label{sec:intro}

\input{introduction}
\vspace{-4mm}
\section{Related work}

\input{related_work}
\vspace{-1mm}
\section{Datasets}

\input{datasets}

\input{tables/librispeech}
\section{Models}
\input{tables/accented}
\input{models}

\vspace{-1mm}
\section{Results}

\input{results}
\vspace{-1mm}
\section{Conclusion}
\input{conclusion}


\bibliographystyle{IEEEbib}
\bibliography{main}

\end{document}

%% file: macros.tex
\newcommand{\rasrconfsmall}{RASR small}
\newcommand{\rasrconfsmalld}{RASR small-distill}
\newcommand{\rasrconfbig}{RASR big}

\newcommand{\sbtransformer}{SB transformer}
\newcommand{\sbcrdnn}{SB crdnn}
\newcommand{\sbwav}{SB wav2vec CV}

\newcommand{\wavvecbasesl}{W2v base 960-960}
\newcommand{\wavveclargesl}{W2v large 960-960}
\newcommand{\wavveclargels}{W2v large 60k-100}
\newcommand{\wavveclargell}{W2v large 60k-960}

\newcommand{\googleapid}{Google API base}
\newcommand{\googleapiv}{Google API video}

\newcommand{\kaldi}{Kaldi}

\newcommand{\werlwc}{$\text{wer}_{l}$}
\newcommand{\weruc}{$\text{wer}_{u}$}
\newcommand{\pninety}{$\text{P}_{90}$ }

\newcommand{\allstar}{ALLSSTAR}

%% file: introduction.tex
The latest state-of-the-art models (SOTA) tend to be trained and evaluated on datasets which are not representative of real-world conditions. 
Public datasets are over indexed on clean read speech by native (US) English speakers, while spontaneous conversations, noisy background, and diverse accents are comparatively rare. 
Besides, most publications today focus on performances on a limited number of datasets without considering other criteria like resilience when being tested out of domain \cite{rasr} or model size efficiency. 
In computer vision some work has already shown that strong bias existed in today’s commercial computer vision systems \cite{gendershades}. 

But what about speech? 
To investigate this issue, we are building a benchmark to confront acoustic models to increasingly difficult conditions representative of real-world settings. 
This way, we hope to get a tool allowing researchers to identify the strengths and weaknesses of their models, and to easily find a method suiting their needs for production systems. 
In this article, we will consider only a handful of models exposed to a limited number of constraints: accented speech, rehearsed speech, and spontaneous conversation. 
We will see however that this intermediate benchmark already gives us some meaningful insights on the following points:

\begin{itemize} 
    \itemsep -1mm
    \item The Librispeech dataset is no longer a representative enough benchmark.
    \item We have noticed a small accuracy gap based on gender.
    \item ASR models show big accuracy variations by accents.
    \item ASR models show very strong performance gap based on the socio-economic status of the speaker.
    \item ASR models show important performance drop on spontaneous speech.
    \item Language models in their current form are not be adapted for spontaneous speech.
\end{itemize}

%% file: related_work.tex
\textbf{Robustness.} Building robust models is becoming a topic of increasing importance for speech. 
In ASR, the best average word error rate (WER) on the Chime Challenge \cite{chime} \cite{chimebest} is about twenty times higher than what is obtained on Librispeech \cite{librispeech} with state-of-the-art models.
Several methods have been developed to increase the robustness of ASR models, RASR for example \cite{rasr} played with smart data augmentation to minimize the performance gap during domain transfer, Ghahremani et al.\cite{lfmmi} focused on multi-task training while Asami et al.\cite{asami} used distillation to improve transfer learning. 

\textbf{ASR benchmark.} Even if Librispeech remains the main reference in speech, there have been several attempts to improve ASR benchmarks.
Recently Aksënova et al, \cite{aksenova-etal-2021-might} surveyed the many potential uses of ASR models in order to define the key features of a representative benchmark.
However, they didn't go as far as testing existing academic models. 
Del Rio et al \cite{earnings21} evaluated both commercial and academic models on real-life data, but they didn't try to identify performance gaps related to accent, gender or social status.

\textbf{Fairness.} As far as we know, there have been only limited work to tackle the issue of fairness in speech technologies. 
Koenecke et al. \cite{koenecke7684} identified significant performance gaps between white and African American speakers, while \cite{kafle} and \cite{guo} pointed the necessity to include data from people with disabilities in current training sets.

%% file: datasets.tex
\subsection{Our reference dataset: Librispeech}

Librispeech \cite{librispeech} is the reference dataset used both for training and testing for many ASR applications. 
It was built using audiobooks from the LibriVox website, filtered and segmented several times in order to get a very clean set of short sequences with an accurate transcript.
Librispeech has a fairly large training set with 960h of labelled audio data, split in two categories \textit{clean} and \textit{other} with their corresponding dev and test sets. 
Today a ``good'' model will often have a WER lower than 6\% even on the \textit{other} test set, going as low as 1.4\% for the best performance on \textit{test-clean}.

\vspace{-2mm}
\subsection{Read speech with accent : \allstar~ and NISP}

\allstar~ \cite{allstar} contains a subset where several speakers read the same text. 
This reduces the likelihood of bias linked to the grammar and vocabulary and allows us to focus exclusively accents.
Some speakers are English natives but most of them are not. 
\allstar~ provides labels for the gender and the mother tongue of each orator. 
The NISP dataset \cite{nisp} contains read recordings from several Indian orators whose native language is one of the following: Hindi, Kannada, Malayalam, Tamil and Telugu.
We focused only on the English section of this dataset.

\vspace{-2mm}
\subsection{Improvised speech with accent: VoxPopuli}

Investigating the rehearsed speech setting, we worked with the VoxPopuli dataset \cite{wang2021voxpopuli} based on recordings from the European Parliament. 
This dataset contains political speech: this means that the orators talk based on a transcript they have usually rehearsed a little bit, but they also improvise and hesitate.
To extract accented English data from the dataset, we used the speaker metadata to target orator who spoke exactly one language on top of English. 
We then assume that these speakers were non-native English speakers and that the other language was their mother tongue. 
This led us to 23 different European accents. 

\vspace{-2mm}
\subsection{Conversational Speech: Buckeye and CORAAL}

Focusing now on conversational speech, we worked with two datasets: Buckeye \cite{buckeye} and CORAAL \cite{coraal}. 
They both have the same setting: a conversation between an two persons. 
Buckeye’s orators are students while CORAAL’s are African Americans from diverse social backgrounds.

%% file: tables/librispeech.tex
\addtolength{\tabcolsep}{-1.5pt} 
\begin{table}[t]
    \centering
    \begin{tabular}{l|c|cc|cc}
    \toprule
      Model & $N_p$& \multicolumn{2}{c|}{Dataset size}&test  & test  \\
            &      & PT & FT& clean &  other\\
      \midrule
      \midrule
      \sbtransformer&255M & - &960h& 3.25&6.79\\
      \sbcrdnn& 202M&- &960h & 2.90 & 8.83\\
      \midrule
      \wavvecbasesl & 95M & 960h & 960h & 3.40 & 8.42 \\
      \wavveclargesl&  317M & 960h & 960h & 2.52 & 5.77 \\
      \wavveclargels & 317M& 60kh & 100h& 2.48 & 5.05 \\
      \wavveclargell & 317M& 60kh & 960h & 2.15 & 4.35 \\
      \midrule
      \rasrconfsmall &28M&- & 4.5kh &3.90 & 8.60 \\
      \rasrconfsmalld &28M& -&4.5kh& 3.80 & 8.40 \\
      \rasrconfbig&317M&-& 4.5kh& 3.20 & 6.40 \\
      \midrule
      \googleapiv &- & - & -& 6.90 & 13.4\\
    \bottomrule
    \end{tabular}
    \caption{
      \textbf{Our reference models} associated with their number of parameters $N_p$, the size of the dataset used at the pretraining stage PT and at the fine-tuning stage FT, and finally the median WER on Librispeech (normalized by speaker).
    \label{tab:librispeech}}
\vspace{-2mm}
\end{table}
\addtolength{\tabcolsep}{1.5pt} 

%% file: tables/accented.tex
\addtolength{\tabcolsep}{-0.5pt} 
\begin{table*}[t]
    \centering
    \begin{tabular}{l|cccccc|cccccc}
    \toprule
      Model  & \multicolumn{6}{c}{\allstar} & \multicolumn{6}{c}{NISP} \\
        &Med. & \pninety & Med. & Med. & Top & Worst & Med.  & \pninety & Med & Med & Best & Worst\\
        & all &  all&  women & men &accent & accent &  all& all  & women & men  & accent & accent  \\
      \midrule
      \midrule
      \sbtransformer &10.4&21.9 & 10.2& 11.0& 7.21 & 20.8 &23.9 &38.6 &21.2 &26.0 & 16.1 & 29.4\\
      \sbcrdnn &13.5&28.5 &12.6& 13.9&8.06  & 25.5  & 36.1&  52.4 & 33.0&38.2 & 25.6 &43.7 \\
      \midrule
      \wavvecbasesl   & 9.50 & 25.5& 9.41& 9.68& 5.70  & 23.8  & 28.9& 43.1& 25.5 & 31.9 & 20.8 & 35.0  \\
      \wavveclargesl   & 6.90 & 15.5& 6.81&7.02 & 4.83 & 13.7 & 15.4 & 26.4& 13.7 & 16.8 & \textbf{10.9} & 19.6\\
      \wavveclargels & 7.34&18.0& 7.34&7.38 & 4.63& 17.1 & 18.9 & 33.3& 16.6 & 19.7 & 12.4 & 23.3 \\
      \wavveclargell  &7.92 & 20.0& 7.92&8.04 & 5.24 & 19.4  & 19.8&33.5 & 17.2 & 21.6 & 14.3 & 21.1 \\
      \midrule
      \rasrconfsmall & 8.45&22.3&7.72 &8.52 & 4.93&21.1 & 16.7 & 28.8& 14.5 & 18.4 & 11.8 & 21.4 \\
      \rasrconfsmalld & 10.3 &21.9&10.1 &10.7 &6.98 & 20.6  &  16.3& 24.6&15.1&17.1 & 12.2 & 20.2 \\
      \rasrconfbig &\textbf{6.45}&\textbf{14.5}&\textbf{5.89} & \textbf{6.70}& \textbf{4.22}& \textbf{13.2} & 14.7& 23.3& \textbf{13.0}& 15.4 & 11.1 & \textbf{18.8}  \\
      \midrule
      \googleapiv  & 8.18 &17.6& 7.66 & 8.35 & 4.91 & 17.3  & \textbf{14.2}& \textbf{22.9}& 13.7 & \textbf{14.9} & 12.0 & \textbf{18.8} \\
    \bottomrule
    \end{tabular}
    \caption{
      \textbf{Median WER, normalized by speaker, on read accented speech}.  For both datasets we consider the median WER and the last decile on the whole set. We also compute the median on the women-only subset and the men-only  subset. Finally we have the lowest and highest median WER obtained on a single accent group.
    \label{tab:allstar}}
\vspace{-2mm}
\end{table*}
\addtolength{\tabcolsep}{0.5pt} 

%% file: models.tex
We explored several open-source methods performing state-of-the-art (SOTA) or near state-of-the-art accuracy on Librispeech.
In order to compare academic models to the performance of an approach closer to production, we added the professional Speech to text Google API, with enhanced \textit{video} setting, to our benchmark.
We wanted to benchmark ready-to-use, end-to-end, speech recognition systems: for this reason we did not train any model on our own and tested each model with both the language model (LM) and the decoding parameters recommended by their authors. 
A first comparison of all of our models with their associated median WER on Librispeech can be found in table \ref{tab:librispeech}.

\vspace{-2mm}
\subsection{SpeechBrain: an accessible ASR toolkit}

SpeechBrain \cite{speechbrain} is a complete open-source speech toolkit implementing a wide range of tasks from speech recognition to speech enhancement. 
Several high quality ASR models are currently available with the Speechbrain toolkit. 
We decided to test two of them: their Transformer model trained on Librispeech and their CRDNN model also trained on Librispeech. 
All models are combined with a Transformer LM \cite{transformerlm} trained on Common Crawl.

\vspace{-2mm}
\subsection{Wav2Vec 2.0 and pretraining}

Wav2vec2.0 \cite{wav2vec} has recently shown that properly pretrained models can bring astounding results with a very limited amount of labelled data: with only 10min of labelled audio a model pretrained without any label can reach a WER as low as 8.2 on Librispeech \textit{test-other} and only 10h are necessary to go below 5. 
Besides, works like XLSR \cite{xlsr} or VoxPopuli \cite{wang2021voxpopuli} proved that pretraining could transfer efficiently across languages. 
For this reason, we hope that pretraining is not only a way to improve performances, but also to deal with robustness. 
We are going to challenge the limits of several wav2vec2.0 models pretrained on audio books and fine-tuned on Librispeech currently available in the fairseq repository. 
Following the recommendations on the fairseq repository, all fairseq models are used with a 4-gram LM trained on Librispeech.

\vspace{-2mm}
\subsection{RASR: increasing ASR's models robustness}

Recently, Likhomanenko and al. \cite{rasr} developed a method to increase the robustness of an ASR model through smart data augmentation strategies.
The idea was to make out-of-domain inference more resilient, especially when dealing with noisy data.
Several RASR models are open-sourced but for the sake of clarity we will only focus on three of them: \rasrconfsmall, \rasrconfbig~  and \rasrconfsmalld. 
The distilled model was obtained by matching the softmax loss distribution of a bigger transformer model: it is a way to compress a big neural network.
For all of theses models we used the a 4-gram LM trained on Common Crawl with the decoding parameters given by Likhomanenko and al. for their release.



%% file: results.tex
\subsection{Metrics}

Tu fully evaluate the robustness and the performances of a model on a given dataset, computing the mean value of the word error rate (WER) is not enough.
Indeed, this value is sensitive to outliers and does not give us any information on the dispersion of the WER distribution.
We therefore decided to consider the median and the last decile \pninety of the distribution.
Furthermore, since the speaker distribution is not always balanced in our test dataset, we always normalize the WER distribution by speaker before estimating any statistic on it.


\vspace{-1mm}
\subsection{Accented speech}

We began with our accented read speech datasets: ALLSSTAR and NISP (see table \ref{tab:allstar}).
As expected, all models showed significantly higher WER when tested on accented data than with Librispeech. 
However, we could observe significant disparities between accents.
Though it isn't mentioned in the table, on the \allstar~ dataset, all models gave their best performances on native English speakers while struggling with non-native speakers from Asian countries. 
We also observed significant disparities in the NISP dataset, with sometimes a two-fold gap between accents.

Finally, although we didn't notice a big difference by gender for any model on the \allstar~ dataset, the gap became wider on NISP, with better performances on women speakers.

\vspace{-1mm}
\subsection{Rehearsed Speech}

From read accented speech we moved to the next step of difficulty: rehearsed speech by non-native English speakers using VoxPopuli (see table \ref{tab:voxpopuli}). 
For each model, we observe a significant performance drop relatively to \allstar.
Besides, even if all models showed significant disparities between accents, these differences were not as strong as the ones observed between Asian and native English speakers in \allstar. 

\input{tables/rehearsed}

\vspace{-1mm}
\subsection{Conversational data}
\input{tables/conversations}

Finally we confronted each model to our 'hardest' decoding setting: conversational speech with Buckeye and CORAAL (see table \ref{tab:conversation}).
On both datasets, we once again, didn’t observe a strong bias along the lines of the speaker’s gender.
On Buckeye, for all models, we could observe a multi-fold increase of the WER compared to Librispeech.
We notice that the Google API, which gave the poorest performance on Librispeech is now the ASR system with the lowest WER.
Meanwhile, all models showed another significant performance drop when tested on CORAAL: in this case, only the big and the distilled RASR models managed to get acceptable error rates.
The speakers in the CORAAL dataset are African-American so the observed performance disparities may well be driven by the speaker’s ethnicity or dialect.
However, what really stroke us was the strong and direct correlation between the socio-economic background of the speaker and the obtained WER: for every model tested in this paper, the wealthier the speaker the more accurate the decoding is.

\vspace{-1mm}
\subsection{Impact of the language model}
\input{tables/LM_impact}

Acoustic models are usually used combined with a LM for a better decoding.
However, LMs are not necessary for the inference and frameworks like fairseq and wav2letter allow us to not use them at all.
For both of them, we ran the inference on all our test datasets without LM, with LM using the decoding parameters provided by their authors, and with the optimal ones giving the best median WER on the target data.
We searched for this parameters on the test data obviously not with the purpose on transferring them for another task but to check if there exists a set of parameters for which the LM brings significantly better results.
We conducted several tests on \wavveclargels and \rasrconfbig (see table \ref{tab:lm_impact}).

As far as \wavveclargels is concerned, the default decoding parameters did not transfer well and even with an optimal setting we couldn't get better than a 10\% improvement.
Indeed, wav2vec models are associated with the Librispeech LM built on audio-books, which is therefore completely out-of-domain when tested on conversational or political speech.
\rasrconfbig \, on the other hand, is coupled with an LM trained on Common Crawl, one of the largest and most diverse open-source datasets available. 
The default decoding parameters transferred rather well on both accented and rehearsed speech, and an optimal setting could bring as much as a 25\% reduction of the WER.
However, they were not adapted for neither of the conversational datasets: we could merely reach a 9\% improvement on Buckeye and a 5\% on CORAAL with optimal decoding.
Therefore, we supposed that standard language models may not be adapted for spontaneous speech: would a pure conversational LM work better?

%% file: tables/rehearsed.tex
\begin{table}[h]
    \centering
    \begin{tabular}{l|ccccc}
    \toprule
      Model & med. & \pninety  & top & worst \\
      & avp & avp & accent & acccent & \\
      \midrule
      \midrule
      \sbtransformer & 22.2 & 30.6 &19.6&28.7 \\
      \sbcrdnn & 28.4 & 40.0 & 24.5 & 39.7\\
      \midrule
      \wavvecbasesl  &   23.3 & 32.5 &19.6 & 30.7  \\
      \wavveclargesl &  18.2  & 25.9& 14.7& 24.0   \\
      \wavveclargels & 18.9 & 28.6&15.6 & 26.8 \\
      \wavveclargell  & 20.4&28.3& 17.9& 27.7  \\
      \midrule
      \rasrconfsmall & 18.6 &29.6 &15.7 & 25.6 \\
      \rasrconfsmalld &  20.2&28.9 &16.9& 26.0  \\
      \rasrconfbig &  \textbf{15.7}&23.5& \textbf{12.3}& 21.9 \\
      \midrule
      \googleapiv & 16.0 & \textbf{23.4} & 12.9 & \textbf{19.7}\\
    \bottomrule
    \end{tabular}
    \caption{
      \textbf{Median WER on accented rehearsed speech} On accented Vox Populi (AVP), the global median WER (med.) is compared to the best median WER by accent and the worst one. 
    \label{tab:voxpopuli}}
\vspace{-5mm}
\end{table}

%% file: tables/conversations.tex
\addtolength{\tabcolsep}{-2pt} 
\begin{table}[h!]
    \centering
    \begin{tabular}{l|cc|cccc}
    \toprule
      Model  & \multicolumn{2}{c|}{Buckeye} & \multicolumn{4}{c}{CORAAL} \\
        & med. & \pninety & med. & \pninety  & \weruc & \werlwc  \\
      \midrule
      \midrule
      \sbtransformer & 36.8& 45.6& 64.9& 85.3&39.3 &70.0\\
      \sbcrdnn& 42.2& 56.8 & 78.3& 97.6& 53.3&81.4 \\
      \midrule
      \wavvecbasesl  & 39.1& 45.3& 67.3& 88.8 & 43.4&77.3 \\
      \wavveclargesl   & 33.6& 42.4 & 78.1&128 & 37.8&96.3\\
      \wavveclargels & 34.6& 49.0& 109& $\infty$ & 40.3&104\\
      \wavveclargell & 34.5& 49.3&81.4 & $\infty$ & 43.9&108 \\
      \midrule
      \rasrconfsmall  & 31.7& 59.6 & 31.7& 60.0 &29.5 & 115\\
      \rasrconfsmalld & 28.4& 44.2& 39.4& 908 & 39.1& 51.1\\
      \rasrconfbig &27.4 &38.2  & \textbf{36.7}& \textbf{51.2}& \textbf{25.2}&\textbf{43.9}\\
      \midrule
      \googleapiv & \textbf{21.9}& \textbf{32.6} & 58.1& $\infty$ & 29.1& 49.7\\
    \bottomrule
    \end{tabular}
    \caption{
      \textbf{Median WER, normalized by speaker, on conversational data}: med. is the median on the whole dataset, \pninety the last decile, \weruc~ and \werlwc~are the median WERs on speakers from the upper and lower working class.
    \label{tab:conversation}}
\vspace{-2mm}
\end{table}
\addtolength{\tabcolsep}{2pt}

%% file: tables/LM_impact.tex

\addtolength{\tabcolsep}{-1.5pt} 
\begin{table}[h]
\centering
\begin{tabular}{l|c|cccc}
\toprule
Model & LM & allstar & vox. &
buck. & coraal \\
\midrule
\midrule
\multirow{ 3}{*}{\shortstack[l]{\rasrconfbig\\ + CommonCrawl LM}}  & no & 8.30 &20.0 &24.9&33.5\\
& std &6.45 &17.1 & 27.4 & 36.7\\
& opt&5.97 &16.6 & 22.6 & 31.8\\
\midrule
\multirow{ 3}{*}{\shortstack[l]{\wavveclargels\\ + LibriSpeech LM}}  & no &7.23 & 18.3& 32.6 & 55.6\\
& std &7.34&18.9 &34.6&109\\
& opt&6.73 & 18.3 & 31.0& 52.2\\
\bottomrule
\end{tabular}
\caption{
      \textbf{Median WER with different decoding parameters}: no language model (no) , default parameters (std), optimal parameters after a random search (opt).
    \label{tab:lm_impact}}
\vspace{-2mm}
\end{table}
\addtolength{\tabcolsep}{1.5pt} 


%% file: conclusion.tex
Even if trained only on clean read speech without any specific effort given to accent adaptation, all wav2vec2.0 and SpeechBrain models transferred quite well when tested on accented data or rehearsed speech. 
The RASR models were the most robust to domain shift overall and could handle conversational speech to some extent, which was not true for the models trained only with read speech. 
This confirms the strength of this training method designed for resilience.
Finally, even though the Google API didn't perform very well on Librispeech, it was consistently robust for both accented and conversational data: this shows that Librispeech is no longer a good reference to evaluate ASR systems.

As far as bias is concerned, none of the models we tested seemed to show a significant performance disparity along gender lines. 
However, we could consistently see a drop in performance when dealing with accented speech, and even worse there seem to be a strong bias related to the speaker’s socio-economic background.